\documentclass[12pt]{article}
\usepackage{epsf,latexsym}
\usepackage{amsmath,amssymb,amsthm,cite}

\theoremstyle{definition}                    

\theoremstyle{remark}

\numberwithin{equation}{section}             

\usepackage[ps,dvips,arc,frame]{xy}

\epsfverbosetrue
\newcommand{\bb}{\begin{eqnarray}}
\newcommand{\ee}{\end{eqnarray}}
\newcommand{\eee}{\nonumber\end{eqnarray}}
\newcommand{\pp}[1]{\begin{pmatrix} #1 \end{pmatrix}}

\newcommand{\rxyz}[2]{{\begin{xy} 0;<2mm,0mm>:<0mm,2mm>::0;0,
,(5,-2)*{a}
,(10,-1.8)*{\bar{a}}
,(15,-2)*{b}
,(20,-2)*{c}
,(25,-2)*{d}
,(30,-2)*{e}
,(35,-1.8)*{\bar{e}}
,(40,-2)*{f}
,(2,-5)*{a}
,(2,-10)*{\bar{a}}
,(2,-15)*{b}
,(2,-20)*{c}
,(2,-25)*{d}
,(2,-30)*{e}
,(2,-35)*{\bar{e}}
,(2,-40)*{f}
,(5,-5)*\cir(#1,0){}
,(10,-5)*\cir(#1,0){}
,(15,-5)*\cir(#1,0){}
,(20,-5)*\cir(#1,0){}
,(25,-5)*\cir(#1,0){}
,(30,-5)*\cir(#1,0){}
,(35,-5)*\cir(#1,0){}
,(40,-5)*\cir(#1,0){}
,(5,-10)*\cir(#1,0){}
,(10,-10)*\cir(#1,0){}
,(15,-10)*\cir(#1,0){}
,(20,-10)*\cir(#1,0){}
,(25,-10)*\cir(#1,0){}
,(30,-10)*\cir(#1,0){}
,(35,-10)*\cir(#1,0){}
,(40,-10)*\cir(#1,0){}
,(5,-15)*\cir(#1,0){}
,(10,-15)*\cir(#1,0){}
,(15,-15)*\cir(#1,0){}
,(20,-15)*\cir(#1,0){}
,(25,-15)*\cir(#1,0){}
,(30,-15)*\cir(#1,0){}
,(35,-15)*\cir(#1,0){}
,(40,-15)*\cir(#1,0){}
,(5,-20)*\cir(#1,0){}
,(10,-20)*\cir(#1,0){}
,(15,-20)*\cir(#1,0){}
,(20,-20)*\cir(#1,0){}
,(25,-20)*\cir(#1,0){}
,(30,-20)*\cir(#1,0){}
,(35,-20)*\cir(#1,0){}
,(40,-20)*\cir(#1,0){}
,(5,-25)*\cir(#1,0){}
,(10,-25)*\cir(#1,0){}
,(15,-25)*\cir(#1,0){}
,(20,-25)*\cir(#1,0){}
,(25,-25)*\cir(#1,0){}
,(30,-25)*\cir(#1,0){}
,(35,-25)*\cir(#1,0){}
,(40,-25)*\cir(#1,0){}
,(5,-30)*\cir(#1,0){}
,(10,-30)*\cir(#1,0){}
,(15,-30)*\cir(#1,0){}
,(20,-30)*\cir(#1,0){}
,(25,-30)*\cir(#1,0){}
,(30,-30)*\cir(#1,0){}
,(35,-30)*\cir(#1,0){}
,(40,-30)*\cir(#1,0){}
,(5,-35)*\cir(#1,0){}
,(10,-35)*\cir(#1,0){}
,(15,-35)*\cir(#1,0){}
,(20,-35)*\cir(#1,0){}
,(25,-35)*\cir(#1,0){}
,(30,-35)*\cir(#1,0){}
,(35,-35)*\cir(#1,0){}
,(40,-35)*\cir(#1,0){}
,(5,-40)*\cir(#1,0){}
,(10,-40)*\cir(#1,0){}
,(15,-40)*\cir(#1,0){}
,(20,-40)*\cir(#1,0){}
,(25,-40)*\cir(#1,0){}
,(30,-40)*\cir(#1,0){}
,(35,-40)*\cir(#1,0){}
,(40,-40)*\cir(#1,0){}
#2\end{xy}}}

\begin{document}

\thispagestyle{empty}

\begin{center}
CENTRE DE PHYSIQUE TH\'EORIQUE \footnote{\, Unit\'e Mixed de
Recherche (UMR) 6207 du CNRS et des Universit\'es Aix-Marseille 1 et 2 \\ \indent \quad \, Sud Toulon-Var, Laboratoire affili\'e \`a la 
FRUMAM (FR 2291)} \\ CNRS--Luminy, Case 907\\ 13288 Marseille Cedex 9\\
FRANCE
\end{center}

\vspace{1.5cm}

\begin{center}
{\Large\textbf{Almost-Commutative Geometries \\ Beyond the Standard Model}} 
\end{center}

\vspace{1.5cm}

\begin{center}
{\large Christoph A. Stephan
\footnote{\, stephan@cpt.univ-mrs.fr} }

\vspace{1.5cm}

{\large\textbf{Abstract}}
\end{center}

In \cite{1,2,3} and \cite{4}  the conjecture is presented that almost-commutative geometries, 
with respect to sensible physical constraints, allow only 
the standard model of particle physics and electro-strong models as Yang-Mills-Higgs theories. In this publication
a counter example will be given.

The corresponding almost-commutative geometry leads to a Yang-Mills-Higgs model which consists of the standard model of particle physics and 
two new fermions of opposite electro-magnetic charge. 
This is the second Yang-Mills-Higgs model within noncommutative geometry, after the standard model, which could be compatible with experiments.
Combined to a hydrogen-like composite particle these new particles 
provide a novel dark matter candidate.

\vspace{1.5cm}

\vskip 1truecm

PACS-92: 11.15 Gauge field theories\\
\indent MSC-91: 81T13 Yang-Mills and other gauge theories

\vskip 1truecm

\noindent September 2005

\newpage

\section{Introduction}

The origin of the internal symmetries of the standard model of particle physics, i.e.
its gauge group $G_{SM}= U(1)_Y \times SU(2)_w \times SU(3)_C$, has been one of the
most prominent mysteries of physics. While the space-time symmetries associated to
general relativity have a clear geometrical interpretation as the diffeomorphisms of the
space-time manifold, such an interpretation was lacking for the standard model. 
A further long standing problem was the unification of these two apparently so 
different  symmetries under one single mathematical roof. This unification is
to be understood in the sense of having
a manifold-like object which exhibits both, internal and space-time symmetries. As a solution to this
unification problem two main propositions have been made. The first and most widely
known is supersymmetry. The second, going back to Heisenberg, inherits the idea
that space-time itself should loose its meaning when high energies are involved.
In the spirit of the uncertainty relation space-time should become {\it noncommutative}
in the sense that the notion of a space-time point looses its meaning, in analogy
with phase-space in quantum mechanics. The original intention was to get rid of
unwanted infinities in quantum field theory. Later it turned out that for classical field
theories these noncommutative spaces could be used to unify the inner and space-time
symmetries since one is no longer restricted to symmetry groups of ordinary manifolds.

In the last few years several approaches to include the idea of noncommutative spaces
into physics have been established. One of the most promising and mathematically 
elaborate is Alain Connes {\it noncommutative geometry} \cite{book} where the main idea is
to translate the usual notions of manifolds and differential calculus into an
algebraic language. Instead of describing manifolds by  atlases, charts
and  differential structures, the geometric information is translated into a set of algebraic
entities which have to meet a number of axioms. These sets are called {\it spectral triples}
and consist, for the calibrating case of a Riemannian spin manifold, of the commutative involution algebra of 
$C^\infty$-functions over the manifold faithfully represented on the Hilbert space of square integrable spinors,
the usual Dirac operator, the charge conjugation operator and a chirality 
operator. These elements comply with the axioms of noncommutative geometry given by Connes and allow to reconstruct all
the geometric data of the manifold, such as dimension, geodesic distance and the notion of a point.
For the main mathematical notions we refer to \cite{book,real,grav} and \cite{costa}.

A crucial feature of this algebraic approach is that the spectral triples and their axioms are independent of
the commutativity of the algebra. It
is in fact possible to construct consistent geometries based on matrix algebras, which exhibit in general a
noncommutative multiplication, if the remaining elements of the spectral triple are chosen in 
an appropriate way. Such an algebra is the direct sum of a finite number of simple matrix
algebras.

Furthermore the tensor product of two spectral triples is again a spectral
triple and produces a consistent geometry. The tensor product of the spectral triple
of a 4-dimensional Riemannian spin manifold and the spectral triple based on a matrix algebra is called
an {\it almost-commutative} spectral triple.

The physical interest of this algebraic description of geometry becomes clear when considering
space-time symmetries and the internal symmetries of the standard model. In noncommutative
geometries the symmetries correspond to the automorphism group of the algebra. It is immediately
clear that the commutative algebra of $C^\infty$-functions over a manifold has the diffeomorphisms of the manifold
as its automorphism group, i.e. the symmetry group of general relativity. On the other hand, 
the gauge group $G_{SM}= U(1)_Y \times SU(2)_w \times SU(3)_C$ of the standard model of particle physics
corresponds
to the automorphism group of the matrix algebra $\mathcal{A}_{SM} = \mathbb{C} \oplus \mathbb{H} \oplus  
M_3(\mathbb{C})$, $\mathbb{H}$ being the quaternions. For this so called {\it internal} or {\it finite}
geometry one takes as Hilbert space the one of the standard model and as the Dirac operator the 
corresponding mass matrix of the fermions.
The tensor product of the spectral triples 
corresponding to these two algebras thus exhibits the combined  space-times and internal symmetries.

Having set up the geometric stage Alain Connes and Ali Chamseddine defined the so called {\it spectral action} 
\cite{cc}.
It is given
by the number of eigenvalues of the Dirac operator, fluctuated with the lifted automorphisms of the
algebra, up to a sufficiently large cutoff.
By use of a technique known as heat-kernel expansion this gives 
as leading terms in the cutoff the desired Einstein-Hilbert Lagrangian plus a cosmological constant and the bosonic
part of the standard model Lagrangian. The Higgs potential appears in the correct form and the Higgs scalar
acquires a geometric interpretation as the gravitational field in the noncommutative part of space-time.
The Lagrangian for the Fermion dynamics and the Yukawa terms are produced by the scalar product on the Hilbert
space, which is given automatically. In combination with the spectral action, any almost-commutative spectral triple 
produces Einsteinian gravity and a Yang-Mills-Higgs theory. 
For a pedagogical introduction see \cite{schuck}.

In this way general relativity and the standard model of particle physics become unified as classical field 
theories. The question that arises immediately is the uniqueness of the choice of the internal algebra, the
corresponding Hilbert space and the Dirac operator. To illuminate the structure of almost-commutative
geometries from a physicists' point of view, they have been classified by imposing a list of physical
requirements for up to four summands in the matrix algebra \cite{1,2,3} and \cite{4}.
The Yang-Mills-Higgs theories were asked to exhibit a non-degenerate spectrum of fermion masses, freedom of
harmful anomalies and a complex representation of the little group on each fermion multiplet.
Furthermore the Hilbert space was asked to be as small as possible which leads to the restriction of 
dealing only with one generation of fermions.
From the necessity to lift the automorphisms of the algebra to the Hilbert space, combined with the
requirement of an anomaly free theory, follow the charges of the fermions with respect to the gauge group
\cite{farewell}.
The essence of this classification is a prominent position of the standard model of particle physics within a 
small set of models, up to an arbitrary number of colours. Apart from a so called {\it electro-strong} model
it is the only Yang-Mills-Higgs theory which fulfils the physical requirements stated above.

The classification can essentially be reduced to a classification of the internal spectral triple, taking
into account only the Yang-Mills-Higgs part of the heat-kernel expansion. Here a diagrammatic
approach is possible \cite{Kraj,pasch}. So called {\it Krajewski diagrams} encode the algebraic data of the
matrix part of the spectral triple, up to the size of the individual summands of the matrix algebra and
the corresponding Hilbert spaces.  Finding the diagrams corresponding to a given number of summands in the
matrix algebra of the spectral triple is a tremendous combinatorial task and thus necessitates the
use of a computer \cite{algo}. For up to four summands  the classification is feasible, since only up to 37
Krajewski diagrams have to be examined. But already for five summands at least 18000 possible 
Krajewski diagrams exist and a complete classification seems out of reach.

The authors of \cite{1,2,3} and \cite{4} have put forward a conjecture stating that the only Yang-Mills-Higgs
models compatible with almost-commutative geometry are either the standard model of particle physics or
electro-strong models. In this paper a counter example will be presented which includes the standard model
as a sub-model and may exhibit interesting phenomenological consequences with respect to dark matter.

\section{The spectral triple}

The Krajewski diagram of the counter example encodes an almost-commutative spectral triple with six summands
in the internal algebra: 

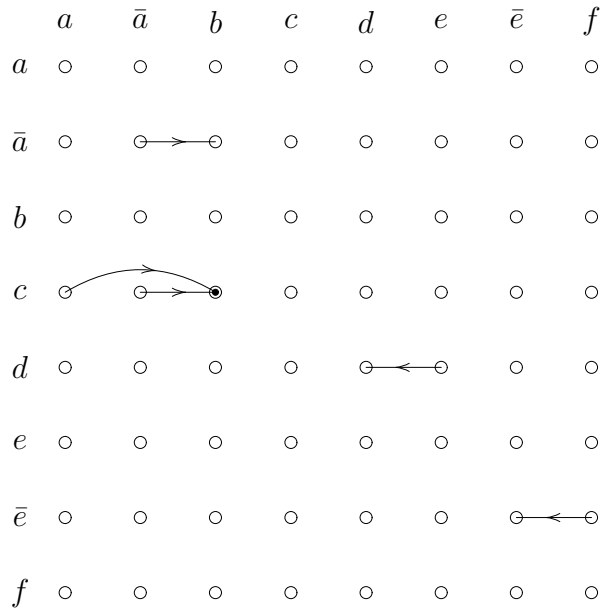
\begin{center}
\begin{tabular}{c}
\rxyz{0.4}{
,(10,-10);(15,-10)**\dir{-}?(.6)*\dir{>}
,(5,-20);(15,-20)**\crv{(10,-17)}?(.6)*\dir{>}
,(10,-20);(15,-20)**\dir{-}?(.6)*\dir{>}
,(15,-20)*\cir(0.2,0){}*\frm{*}
,(30,-25);(25,-25)**\dir{-}?(.6)*\dir{>}
,(40,-35);(35,-35)**\dir{-}?(.6)*\dir{>}
} \\
\\
\end{tabular}
\end{center}
One can clearly see the sub-diagram of the standard model in the upper $4\times 4$ corner. Essentially the diagram is a
direct sum of the diagrams (1) and (17) given in \cite{1}. One can find all the necessary translation rules between Krajewski
diagrams and the corresponding spectral triples in \cite{Kraj} and \cite{1}. The matrix is already {\it blown up} in the sense that the
representations of the complex parts of the matrix algebra 

\bb
\mathcal{A}= \mathbb{C} \oplus \mathbb{H} \oplus M_3(\mathbb{C}) \oplus \mathbb{C} \oplus \mathbb{C} \oplus \mathbb{C} \ni (a,b,c,d,e,f)
\nonumber
\ee
are fixed:
\bb
\rho_L (a,b,c,d,e,f)&=&  \pp{ b \otimes 1_3 & 0 & 0 & 0 \\ 0 & b & 0 & 0 \\ 0 & 0 & d & 0 \\ 0& 0& 0& \bar{e} },
\quad
\rho_R (a,b,c,d,e,f)=  \pp{ a 1_3 & 0 & 0 & 0 & 0 \\ 0 & \bar{a} 1_3 & 0 & 0 & 0\\ 0 & 0 & \bar{a}  & 0 & 0 \\ 0& 0& 0& e & 0 \\
0&0&0&0&f }, 
\nonumber \\ \nonumber
\ee

\bb
\rho_L^c (a,b,c,d,e,f)&=&  \pp{  1_2 \otimes c & 0 & 0 & 0 \\ 0 & \bar{a} 1_2 & 0 & 0 \\ 0 & 0 & d & 0 \\ 0& 0& 0& e },
\quad
\rho_R^c (a,b,c,d,e,f)=  \pp{ c & 0 & 0 & 0 & 0 \\ 0 & c & 0 & 0 & 0\\ 0 & 0 & \bar{a}  & 0 & 0 \\ 0& 0& 0& e & 0 \\
0&0&0&0& \bar{e}} 
\nonumber
\ee
These representations are faithful on the Hilbert space given below and serve as well to construct the lift of the automorphism group.
As pointed out in \cite{farewell} the commutative sub-algebras of $\mathcal{A}$ which are equivalent to the complex numbers serve 
as receptacles for the $U(1)$ subgroups embedded in the automorphism group $U(3)$ of the $M_3(\mathbb{C})$ matrix algebra. In
this way the standard model gauge groups $U(1)_Y$ and $SU(3)_C$ are generated. Roughly spoken each diagonal entry of the
representation of the algebra can be associated to fermion multiplet. For example the first entry of $\rho_L$, $b \otimes 1_3$,
is the representation of the algebra on the up and down quark doublet, where each quark is again a colour triplet.

The Hilbert space is given by the direct sum of the standard model Hilbert space, for details see \cite{schuck}, and a
Hilbert space containing two new fermions

\bb
\mathcal{H} = \mathcal{H}_{SM} \oplus \mathcal{H}_{new},
\nonumber
\ee
where
\bb
\mathcal{H}_{new} \ni \pp{\psi_{1L} \\ \psi_{2L}} \oplus \pp{\psi_{1R} \\ \psi_{2R}} \oplus \pp{\psi_{1L}^c \\ \psi_{2L}^c} \oplus 
\pp{\psi_{1R}^c \\ \psi_{2R}^c}.
\nonumber
\ee

The wave functions $\psi_{1L}$, $\psi_{2L}$, $\psi_{1R}$ and $\psi_{2R}$ are the respective left and right handed Dirac 4-spinors. 
The initial internal Dirac operator,  which is to be 
fluctuated with the lifted automorphisms is chosen to be the mass matrix

\bb
\mathcal{M}= \pp{ \pp{m_u & 0 \\ 0 & m_d} \otimes 1_3 &0&0&0 &0\\ 0&0&m_e&0&0 \\ 0&0&0&m_1&0 \\ 0&0&0&0&m_2 },
\nonumber
\ee
with $m_u,m_d,m_e,m_1,m_2 \in \mathbb{C}$. 

It should be pointed out that the above choice of the summands of the matrix algebra, the Hilbert space and the Dirac operator
is rather unique, if one 
requires the Hilbert space to be minimal and the fermion masses to be non-degenerate.
Fluctuating the Dirac operator and calculating the spectral action gives the usual Einstein-Hilbert action, the Yang-Mills-Higgs
action of the standard model and a new part in the Lagrangian for the two new fermions:

\bb
\mathcal{L}_{new} &=& i \psi_{1 L}^\ast D_1 \psi_{1 L} + i \psi_{1 R}^\ast D_1 \psi_{1 R} + m_1 \psi_{1 L}^\ast \psi_{1 R} +
 m_1 \psi_{1 R}^\ast \psi_{1 L}
\nonumber  \\
&&+ \; i \psi_{2 L}^\ast D_2 \psi_{2 L} + i \psi_{2 R}^\ast D_2 \psi_{2 R} + m_2 \psi_{2 L}^\ast \psi_{2 R} +
m_2 \psi_{2 R}^\ast \psi_{2 L}.
\nonumber
\ee

The covariant derivative turns out to couple the two fermions to the $U(1)_Y$ sub-group of the standard model gauge group,

\bb
D_{1/2} &=& \partial + \frac{i}{2} \, g' \, Y_{1/2} B 
\nonumber \\
&=& \partial + \frac{i}{2} \, e \, Y_{1/2} A - \frac{i}{2} \, g' \, \sin \theta_w Y_{1/2} Z,
\nonumber
\ee
where $B$ is the gauge field corresponding to $U(1)_Y$, $A$ and $Z$ are the photon and the Z-boson fields, e is the
electro-magnetic coupling and $\theta_w$ is the weak angle. The hyper-charge
$Y_{1/2}$ of the new fermions can be any non-zero fractional number with $Y_1 = - Y_2$ so that $\psi_1$ and $\psi_2$ have
opposite electrical charge. $Y_1=0$ is not allowed since this would conflict with the postulated
complex representation of the little group on the fermions. For simplicity $Y_1=1$ may be chosen which results in opposite
electro-magnetic charges $\pm e$ for $\psi_1$ and $\psi_2$.

The masses of the standard model fermions are obtained by minimising the Higgs potential. It turns
out that the masses $m_1$ and $m_2$ of the new fermions do not feel the fluctuations of the Dirac operator and are thus 
Dirac masses which do not stem from the Higgs mechanism. This is due to the vector like coupling of the gauge group induced
by the lift of the automorphisms. Consequently the Dirac masses do not break gauge invariance.

\section{Speculations on dark matter}

The question which arises is whether these new particles could be realised in nature. It is tempting to identify them
with the missing dark matter, but the apparent electro-magnetic charge seems to pose a problem. One could though speculate
that dark matter does not consist of elementary particles but instead of hydrogen-like composite particles built out of
the oppositely charged new fermions. Assuming that the mass $m_1$ of the lighter particle is of the order of 500 GeV or higher
and $m_2 \geq m_1$,  
the composite two particle system would have a radius comparable to a neutron or even smaller. This estimation of the
mass has been chosen such that no conflict with current accelerator experiments should occur.

The ionisation energy of the two-particle system would be of the
order of a few MeV, comparable to gamma ray energies of radioactive nuclei. If one assumes further a dark matter density
of $\rho_{DM} \sim 0.5 {\rm TeV}/m^3$ in the galactic plane this would result in a particle density of $\rho_{part}\leq 0.5 $ particles$/m^3$.
These composite particles would freeze out in the early universe when the temperature drops below the ionisation energy, in
analogy with the freeze out of Hydrogen in the period of recombination.

Since these hydrogen-like particles are electro-magnetically neutral and do not interact strongly it should be very difficult to
detect them directly. It remains of course to investigate whether this kind of dark matter is compatible with the established cosmological
models, especially baryogenesis and whether it is already possible to exclude them with current experiments.

\section{Conclusions}

A counter example to the conjecture that almost-commutative geometries, with respect to sensible physical constraints, allow only 
the standard model of particle physics and electro-strong models as Yang-Mills-Higgs theories, given in \cite{1,2,3} and \cite{4},
has been presented. The corresponding Yang-Mills-Higgs model
predicts two new fermions with opposite electro-magnetic charge and weak interaction through the Z-boson. Their masses are not
acquired via the Higgs mechanism but are Dirac masses, which can be chosen arbitrarily. Combining these particles into 
a hydrogen-like composite 
particle could provide a new candidate for dark matter, if compatible with current experiments. Further investigations of these
composite particles are clearly necessary. It should be pointed out that the mathematical requirements on spectral triples are
extremely tight and that the physical requirements presented above are most general. It is thus highly non-trivial that Yang-Mills-Higgs
models containing the standard model of particle physics exist at all in the context of almost-commutative geometry.

\vskip1cm
\noindent
{\bf Acknowledgements:} The author would like to thank T. Sch\"ucker for his advice and support and J.H. Jureit and R. Oeckl
for helpful suggestions and careful proofreading.

\end{document}